\documentclass[aps,prl,showpacs,twocolumn,floatfix,superscriptaddress]{revtex4}
\usepackage{graphicx}
\usepackage{amsfonts}
\usepackage{amssymb}

\begin{document}

\title{Slowing heavy, ground-state molecules using an alternating gradient decelerator}

\author{M. R. Tarbutt}
\affiliation{Blackett Laboratory, Imperial College, London SW7
2BW, UK}

\author{H. L. Bethlem}
\affiliation{FOM-Institute for Plasma Physics Rijnhuizen, P.O. Box
1207, NL-3430 BE Nieuwegein, The Netherlands}

\author{J. J. Hudson}
\affiliation{Blackett Laboratory, Imperial College, London SW7
2BW, UK}

\author{V. L. Ryabov}
\affiliation{Petersburg Nuclear Physics Institute, Gatchina,
Leningrad 188300, Russia}

\author{V. A. Ryzhov}
\affiliation{Petersburg Nuclear Physics Institute, Gatchina,
Leningrad 188300, Russia}

\author{B. E. Sauer}
\affiliation{Blackett Laboratory, Imperial College, London SW7
2BW, UK}

\author{G. Meijer}
\affiliation{Fritz-Haber-Institut der Max-Planck-Gesellschaft,
Faradayweg 4-6, D-14195 Berlin, Germany}

\author{E. A. Hinds}
\affiliation{Blackett Laboratory, Imperial College, London SW7
2BW, UK}

\date{\today}


\begin{abstract}
Cold supersonic beams of molecules can be slowed down using a
switched sequence of electrostatic field gradients.  The energy to
be removed is proportional to the mass of the molecules. Here we
report deceleration of $^{174}$YbF, which is 7 times heavier than
any molecule previously decelerated. We use an alternating
gradient structure to decelerate and focus the molecules in their
ground state. We show that the decelerator exhibits the axial and
transverse stability required to bring these molecules to rest.
Our work significantly extends the range of molecules amenable to
this powerful method of cooling and trapping.
\end{abstract}

\maketitle

It has recently become possible to decelerate cold supersonic
beams of light polar molecules in suitable excited states using
electrostatic forces. This has been demonstrated with beams of CO
\cite{Bethlem(1)99}, ND$_3$ \cite{Bethlem(1)00,Bethlem(1)02}, and
OH \cite{Bochinski(1)03}. Once the molecules are sufficiently slow
they can be trapped electrostatically
\cite{Bethlem(1)02,Crompvoets(1)01}.  The key feature of an
electrostatic decelerator is that molecules move through the
fringe field of a capacitor from low to high electric field. If
the molecules are weak-field-seekers the increase in their
potential energy causes them to slow down. The field is then
switched off and the molecules fly out of the capacitor without
being accelerated. With many capacitors along the beam line, this
process can be repeated until the molecules are brought to rest.
The highest electric fields are on the surfaces of the electrodes,
resulting in a transverse force towards the beam axis that focuses
the weak-field-seeking molecules as they propagate through the
decelerator \cite{Bethlem(1)02}.

Until now, no ground state beam has been decelerated because
ground state molecules are strong-field seekers.
Strong-field-seekers can be decelerated by reversing the charging
sequence so that the field is off when the molecules enter each
capacitor and switched on before they leave. However, Maxwell's
equations prevent the simple focussing that weak-field seekers
enjoy because it is not possible to have a static maximum of
electric field in free space \cite{Wing(1)84}. This difficulty can
be overcome by the method of alternating gradient focussing,
whereby a sequence of alternating focussing and de-focussing
lenses results in a net focussing effect. Recently a beam of
strong-field-seeking CO molecules in the metastable a$^3\Pi$ state
was slowed from 275\,m/s to 260\,m/s in such a decelerator
\cite{Bethlem(2)02}. Here we report the deceleration and focussing
of a supersonic $^{174}$YbF beam in the ground state
$X^2\Sigma^+(v=0,N=0)$ and we show that it is feasible to bring
such molecules to rest. These are seven times heavier than any
molecules previously decelerated, so the amount of kinetic energy
to be removed is proportionally larger. If the number of
decelerator stages is to remain practical each stage must remove a
large amount of energy. Much more energy can be removed from heavy
molecules in their ground state than in any weak-field-seeking
state.

\begin{figure}[tbp]
\begin{center}
\includegraphics[width=3.6in]{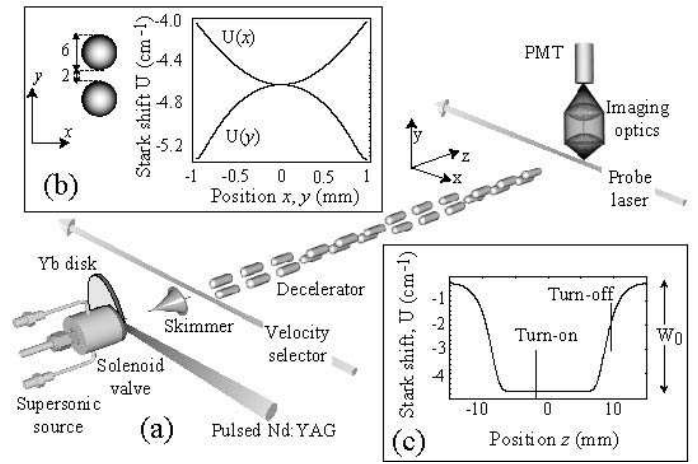}
\caption{\label{ExperimentSetup}(a) The beam line. (b) Left: end
view of one decelerator stage. Graph: Stark shift of ground state
YbF within the stage versus {\textit x} and {\textit y} for
electrode voltages of $\pm 10$\,kV. (c) Same Stark shift versus
{\textit z}, plotted for one stage.}
\end{center}
\end{figure}

The ability to decelerate heavy molecules and ground state
molecules greatly extends the range of species amenable to this
powerful method of cooling and trapping. YbF is of particular
interest because it offers one of the most sensitive ways to
search for elementary particle physics beyond the standard model
through a measurement of the electric dipole moment of the
electron \cite{Hudson(1)02}. This important quantity can be
measured more sensitively by slowing the molecules. Furthermore,
deceleration would enhance experiments that study possible
connections between the homochirality of biomolecules and parity
violation \cite{Daussy(1)99, Ziskind(1)02} and would improve the
sensitivity of nuclear anapole measurements \cite{DeMille(1)03}.

Our beam line is shown in Fig.\,\ref{ExperimentSetup}(a). The
supersonic source uses 4 bar of Ar, Kr or Xe mixed with a small
fraction of SF$_{6}$ as the carrier gas. A solenoid valve releases
this gas every 0.2\,s in 100\,$\mu$s pulses that expand through a
nozzle into a vacuum of $\sim 5\times10^{-5}$\,mbar. A disk of Yb
is positioned immediately outside the nozzle. The edge of this
disk is ablated by the unfocussed 1064\,nm output of a 10\,ns,
20\,mJ pulsed Nd:YAG laser. The Yb and SF$_{6}$ react to form YbF,
which becomes entrained in the gas pulse and thermalizes with it.
The molecules then pass through a 1\,mm-diameter skimmer, 82\,mm
from the nozzle, into a high vacuum. This source produces YbF
beams as slow as 290\,m/s with translational and rotational
temperatures as low as 1.4\,K \cite{Tarbutt(1)02}. A 552\,nm laser
beam, the velocity selector, crosses the molecular beam 108\,mm
downstream from the skimmer. This laser beam pumps molecules out
of the ground state except during an adjustable period of
typically 10-20\,$\mu$s when it is switched off. The moment at
which a molecule crosses the laser beam is mainly determined by
its speed. Consequently, the resulting pulse of ground state
molecules is a velocity group whose width and center are
determined by the timing of the laser pulse. A further 4\,mm
downstream the molecules enter the decelerator. This has 12
stages, each comprising a pair of 20\,mm-long stainless steel rods
parallel to the beam axis {\it z} and rounded to have
hemispherical ends. The rods are 6\,mm in diameter, with a 2\,mm
gap between them, as indicated on the left of
fig.\,\ref{ExperimentSetup}(b). The stages are separated from each
other by 10\,mm gaps. They are arranged in four groups of three,
alternating between vertical ($y$) and horizontal ($x$)
orientation, as illustrated in fig.\,\ref{ExperimentSetup}(a).
Finally, a 552\,nm probe laser, 258\,mm from the end of the
decelerator and perpendicular to the molecular beam, selectively
excites one hyperfine sub-level of the ground state. The
time-evolution of the fluorescence is detected by a
photomultiplier.

The decelerator electrodes are operated at $\pm 10$\,kV, giving an
electric field on the axis of nearly 100\,kV/cm at the center of
each stage. Fig.\,\ref{ExperimentSetup}(b) shows the resulting
Stark shift of the molecules in a vertical stage versus transverse
position $x$ or $y$.  This potential focusses the molecules toward
the beam axis in the $x$-direction and defocusses then along $y$.
It is approximately harmonic over most of the region between the
electrodes. The transverse motion can therefore be described using
transfer matrices (as in optics), and the transverse oscillations
are stable provided $-2<$Tr(M)$<2$, where M is the transfer matrix
for one unit of the periodic structure \cite{Bethlem(2)02}. We
have chosen to alternate the focussing and defocussing axes after
every third lens, rather than between each lens. In this way, we
are able to maximize the number of deceleration stages per unit
length and still satisfy the alternating gradient stability
criterion.

Fig.\,\ref{ExperimentSetup}(c) shows the variation of the Stark
shift along the beam axis through one stage. As the beam pulse
passes through the stage the voltages are switched on and off by
fast high-voltage switches under the control of a 50\,MHz pattern
generator. Fig.\,\ref{ExperimentSetup}(c) illustrates where the
field might be switched on and off to decelerate a molecule moving
to the right. The switching sequence that decelerates a pulse of
molecules through many stages is designed so that a particular
molecule, known as the synchronous molecule, reaches the same
positions within each stage at the moments of switching on and
off. The switch-off position is described by a phase angle $\phi$
defined through the energy loss of the synchronous molecule per
stage $W_{0} \sin \phi$, where $W_{0}$ is the maximum possible
energy loss per stage. By choosing $\phi < 90^{\circ}$ we ensure
that those molecules running ahead of the synchronous molecule are
decelerated more, while those lagging behind are decelerated less.
This phase stability is important for a decelerator because it
allows a finite velocity group of molecules to propagate as a
bunch \cite{Bethlem(2)00}.

\begin{figure}[tbp]
\begin{center}
\includegraphics[height=2.4in]{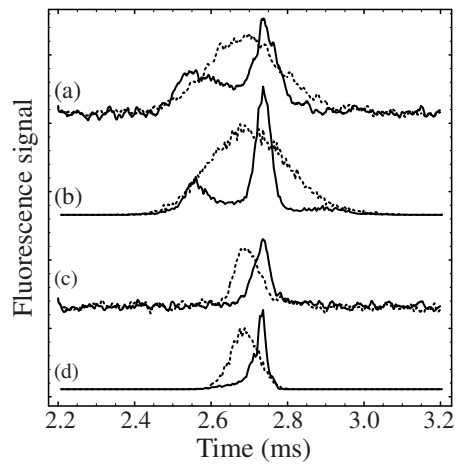}
\caption{\label{Decel300}Time-of-flight profiles for 298\,m/s YbF
beams with decelerator off (dotted lines), and on (solid lines).
The time origin is defined by the Nd:YAG pulse. (a) Experiment
using the whole pulse of molecules. (b) Simulation for (a). (c)
Experiment using a velocity-selected pulse of molecules. (d)
Simulation for (c).  The vertical axes of the simulations are
scaled to account for transverse focussing which is not otherwise
included.}
\end{center}
\end{figure}

\begin{figure*}[btp]
\begin{center}
\includegraphics[width=7in]{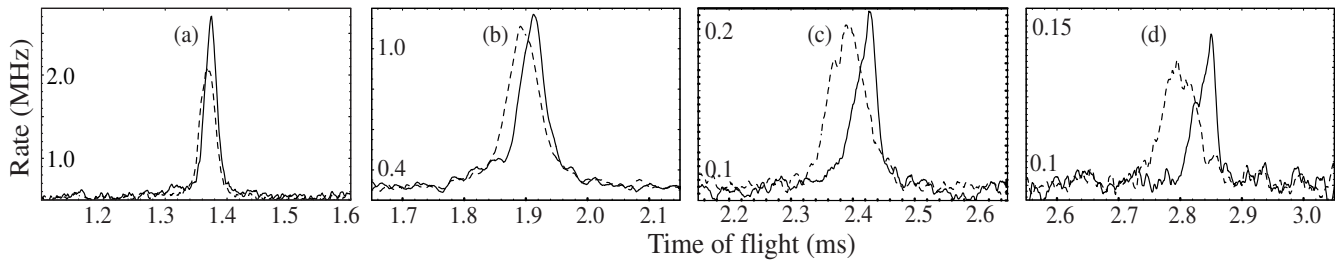}
\caption{\label{DecelSeries}Deceleration of YbF beams from various
initial speeds. Dotted (solid) lines show time-of-flight profile
with decelerator off (on). The y-axes indicate the fluorescence
rates in the detector. (a) YbF in Ar. $v_{i}=586.3 \pm 1.0$\,m/s,
$\Delta v=5.1 \pm 0.9$\,m/s, $\Delta E=48.1 \pm 8.7$\,cm$^{-1}$.
(b) YbF in Kr. $v_{i}=421.8 \pm 0.7$\,m/s, $\Delta v=6.9 \pm
0.7$\,m/s, $\Delta E=46.8 \pm 4.8$\,cm$^{-1}$. (c) YbF in Xe.
$v_{i}=335.1 \pm 0.5$\,m/s, $\Delta v=9.3 \pm 0.5$\,m/s, $\Delta
E=50.0 \pm 2.4$\,cm$^{-1}$. (d) YbF in Xe cooled to 198\,K.
$v_{i}=286.5 \pm 0.5$\,m/s, $\Delta v=10.2 \pm 0.4$\,m/s, $\Delta
E=46.6 \pm 2.1$\,cm$^{-1}$.}
\end{center}
\end{figure*}

The dotted curve in fig.\,\ref{Decel300}(a) shows the YbF
time-of-flight distribution with the decelerator off, using Xe
carrier gas pre-cooled to 210\,K. We deduce from this distribution
that the YbF pulse has a mean velocity of 298\,m/s distributed
with a full width at half maximum (FWHM) of 26\,m/s, corresponding
to a temperature of 2.7\,K. When the decelerator is turned on
(solid line) at a phase angle of $60^{\circ}$, approximately half
the molecules gather into a narrow bunch whose speed is 287\,m/s
with 13\,m/s FWHM. Most of the molecules inside the bunch have
been decelerated, although some of the slowest molecules from the
original distribution are accelerated into the bunch, as indicated
by the reduction in the number of molecules at long arrival times
when the decelerator is turned on. The bunch will stay together as
more stages of deceleration are added to bring it to rest.

To test our understanding of the decelerator, we simulate the
motion of molecules through the apparatus by integrating the axial
equation of motion using the calculated potential on the axis of
the beam line, which is plotted in Fig.\,\ref{ExperimentSetup}(c).
For the initial velocity distribution in the simulation, we take a
beam whose temperature and central velocity have the values
derived from the dotted data in fig.\,\ref{Decel300}(a). The
result of this calculation, plotted in fig.\,\ref{Decel300}(b), is
qualitatively very similar to the experimental result, indicating
that the motion of the molecules is indeed roughly separable into
axial and transverse components and showing that we understand the
basic deceleration process.

Fig.\,\ref{Decel300}(c) shows the results obtained with the
velocity selector turned on and timed to select the central $20
\mu s$ section of the pulse. The resulting velocity distribution
remains centered on 298\,m/s but now has a FWHM of only 8.5\,m/s.
When the decelerator is turned on, this entire distribution is
shifted down in velocity by 10\,m/s and the time-of-flight profile
is narrowed to 70\% of its original width because the decelerator
slows down the faster molecules in the bunch more than the slower
ones. Fig.\,\ref{Decel300}(d) confirms that these results are also
well described by a simple simulation of the axial motion.

We have further tested the decelerator using four other initial
speeds, chosen by varying the mass and temperature of the carrier
gas, with the results shown in fig.\,\ref{DecelSeries}. In each
experiment the velocity selector was used to narrow the velocity
distribution to a FWHM in the range 9-14\,m/s. This longitudinal
velocity range lies within the phase stability range of the
decelerator, meaning that the entire velocity-selected pulse can
be decelerated. In all four experiments, the phase angle is set to
$60^{\circ}$, corresponding to a total energy loss of
$46$\,cm$^{-1}$. The measured energy loss confirms this, showing
that the decelerator works correctly over a wide range of initial
speeds. All the beams become 20-30\% narrower in their
time-of-flight profile when the decelerator is switched on as a
result of the phase stability. The slowest beam we have produced
is shown in Fig.\,\ref{DecelSeries}(d). Here the speed of the
molecules entering the decelerator is only 286.5\,m/s and the
decelerator reduces their energy by 7\%.

\begin{figure}[tbp]
\begin{center}
\includegraphics[width=2.5in]{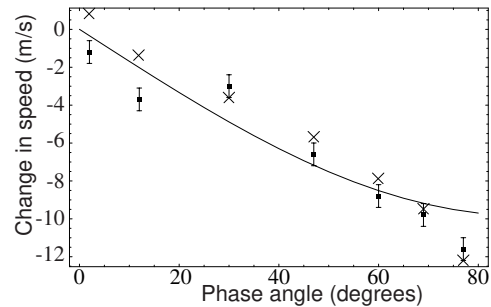}
\caption{\label{speedVsPhase}Change in speed of molecules versus
decelerator phase angle for a beam of initial speed 335\,m/s.
Points: measured data. Solid Line: Simple theory for synchronous
molecule. Crosses: Numerical simulation.}
\end{center}
\end{figure}

The final speed of the molecules can be adjusted by making an
appropriate choice of phase $\phi$ in the high voltage switching
pattern, as shown by the data points in figure \ref{speedVsPhase}.
Here we have simply taken the peak of the decelerated pulse as a
measure of the speed. The solid line shows the expected change for
a synchronous molecule, corresponding to an energy loss of $W_0
\sin \phi$ per stage. Crosses show the result of a numerical
simulation that takes into account the entire ensemble of
molecules. The uncertainty in these simulations is the same as the
error bars on the data. As the phase is increased toward
$90^{\circ}$, the measured change of speed shows reasonable
agreement with the theories. Figures \ref{Decel300},
\ref{DecelSeries} and \ref{speedVsPhase} show that this
alternating gradient decelerator works well over a wide range of
operating parameters and that its deceleration and phase stability
properties are well understood.

\begin{figure}[tbp]
\begin{center}
\includegraphics[width=2.6in]{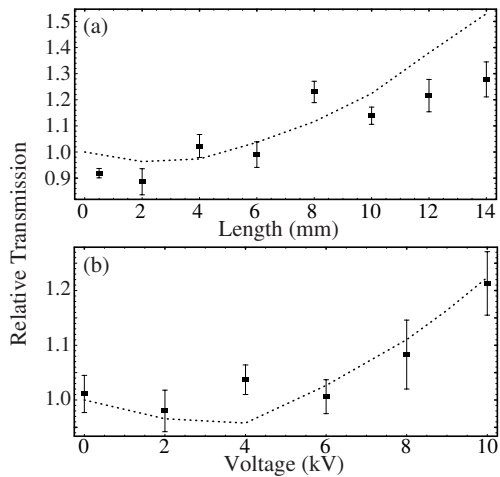}
\caption{\label{FocusPlot}Normalized decelerator transmission
versus (a) effective lens length at a voltage of $\pm$10\,kV, and
(b) electrode voltage for a fixed lens length of 10\,mm. Dotted
lines: calculation using matrix approach.}
\end{center}
\end{figure}

Transverse confinement is important for the successful operation
of the decelerator. In order to study this experimentally we
measured the number of molecules reaching the detector for a range
of electrode voltages and lens lengths. Although the physical
length of the electrodes is fixed, we can vary the length of time
for which each lens acts on the molecules by changing the turn-on
point in the timing sequence (fig.\,\ref{ExperimentSetup}(c)),
keeping the turn-off point fixed. Measurements on a 590\,m/s beam
are shown in Fig.\,\ref{FocusPlot}, where we plot the total number
of molecules transmitted through the decelerator, normalized to
the number when the decelerator is off for various lens lengths
(\ref{FocusPlot}(a)) and various electrode voltages
(\ref{FocusPlot}(b)). The dotted lines show the expected
transmission of molecules through our decelerator in the absence
of edge effects, lens aberrations and misalignments. This was
calculated using transfer matrices, sampling over a realistic
range of initial transverse velocities and positions. At low lens
power, the calculated transmission dips below unity because the
defocussing lenses are 10\% stronger than the focussing ones. At
higher power, alternating gradient focussing increases the
transmission. Our data follow the expected trend, whereas the only
previous study of alternating gradient deceleration of molecules
showed a 20-fold shortfall in the transmission, a problem
attributed to misalignments of the electrodes \cite{Bethlem(2)02}.
By varying the phase angle $\phi$ we can control the fraction of
time that the molecules spend in the fringe fields of the lenses
and we find that this alters the focussing. We believe that the
fringe fields, not accounted for in the matrix model, are the
cause of the small discrepancy between theory and experiment
observed in fig.\,\ref{FocusPlot}(a). This observation underlines
the importance of keeping the fringe fields under review while
designing alternating gradient decelerators.

We have demonstrated that heavy ground state molecules can be
slowed down by using an alternating gradient decelerator. The
technique we have demonstrated can be applied to a wide range of
heavy polar molecules. We have used a 12-stage decelerator
operated at a field of 100\,kV/cm with a deceleration phase angle
of $60^{\circ}$ to remove 7\% of the kinetic energy of YbF
molecules. We were able to raise the electric field to 140\,kV/cm
without electric field breakdown. At this higher field, 110 stages
would bring the molecules to rest. They could then be coupled into
an electrodynamic trap \cite{Shimizu(1)92, Peik(1)99} or a wire
trap \cite{Sekatskii(1)96}, or they could be switched into a
weak-field seeking state and stored in an electrostatic trap e.g.
a quadrupole trap \cite{Wing(1)80} or chain-link trap
\cite{Shafer-Ray(1)03}. Trapping heavy molecules will provide a
spectacular increase in the spectroscopic resolution available,
leading to new applications in metrology, quantum chemistry and
fundamental physics.

We are indebted to Henrik Haak and Andr\'{e} van Roij for
mechanical design work and to Victor Ezhov for valuable
discussions. This work was supported in the UK by PPARC, EPSRC and
 JIF, by the EU `Cold Molecules' network, and in Russia by RFBR grant 02-02-17090.
  HLB acknowledges a Talent fellowship from the NWO.

\end{document}